**Brief Communication Arising from a Corrigendum to "Atomic-resolution chemical analysis using a scanning transmission electron microscope"**

John Silcox, David A. Muller

School of Applied and Engineering Physics, Cornell University, Ithaca, NY

In a recent corrigendum[1] the authors of a 1993 Nature paper[2] admitted to misleading the referees and editors of Nature as to the origins and processing of their data. During the review process, new data were substituted for only part of the initial data, an action that was not remedied during the final stages of acceptance and publication. In itself, this action completely undercuts the primary claim of the paper, makes it inappropriate for an archival record and gives firm grounds for retraction of the article. The assertion that the proffered correction '*in no way affects the key scientific claim of the paper*' is completely wrong. Modern computational methods make it much easier to analyze data than at that time and we have used these extensively to make this point in the supplementary material for readers to evaluate for themselves. We summarize key points below.

Our central objection, then and now, is that the original data did not show an atomically abrupt interface. No amount of data processing can turn a physically rough interface into a sharp one, and while the underlying elastic image and spectral data of Fig 3 showed a diffuse, rough interface, the derived line profile of Fig 4b appeared unphysically abrupt. The authors' response to the initial criticism of the referee was not to defend analysis of the data, but rather to substitute a new image and different spectra (two of which were duplicates) for three of those originally submitted. In particular, spectrum #5 (see Suppl. Figs. **1** and **2**) with a small but undeniable cobalt edge was replaced by a spectrum with no evidence at all of a cobalt edge. Despite these changes, in the 2[nd] response to the referee the authors claimed that the data was unchanged, and had instead been replotted to show the jump in the cobalt spectrum across the interface more clearly. Only by reading part (b) of the new captions for Fig. 3 does it become clear that the authors claimed to have reanalyzed the data using a new method of background subtraction. The caption further explains in some detail that this background was applied to both sides of the interface (i.e. including curves 1-4). It is at this point that JS as referee pointed out "spectra 1-4 are identical with the previous set" and "only spectra 5, 6 and 7 have been treated removing the residual data noted earlier at the Co L2,3 edge position". The authors specifically denied this was the case, but in the recent corrigendum now admit that this is exactly what they did.

Whether spectra 5-7 in the final and published version are the same spectra as in the 1[st] version (and conference proceeding[3]) differing only in the method of background subtraction is the critical question. The final version (5'-7') was supposed to have used a scalar multiple of the same reference curve subtracted from each spectrum. If this were indeed the case, then subtracting spectra 5 from 5', 6 from 6' and 7 from 7' should leave as a residual only a scaled version of the reference curve sitting on a smooth background. We have scanned and digitized the two data sets (Suppl Figs. **3** and **4**). Plotting one against the other shows a good 1:1 correspondence, demonstrating that they have been correctly scaled (Supp. Fig **5**.) Finally we perform the subtraction on all 7 spectra (Supp. Fig **6**). The small, featureless residual for curves 1-4 give the magnitude of the errors of our digitization process, showing that these curves are essentially the same. However, the large and differing residuals for curve 5'-7' show that they could not have been processed in the manner claimed in the corrigendum, or if they were, then they were in no way related to the original curves 5-7. In other words, the spectra that are supposed to show the atomically abrupt step down appear to have been replaced by data that has no connection to the original data set (Supp. Fig. **2**). Note in particular, the large dip in curve 5'-5 (Supp. Fig **6**) between 4 and 8 on the x-axis, due to the missing Co L3 peak in the 'new' data. Any further processing of this data (as done in Nature Fig 4) is pointless given its dubious origins. This alone is grounds for retraction.

Turning now to the image substitution, Supp. Fig 7 shows the image in question, Fig 4b of the MSA proceeding[3] which was also the same image shown in the first draft of the Nature paper. In brief, this image is a map of the elastic scattering recorded point by point as the electron probe is scanned across the sample. To assess the visual characteristics of the image, Suppl. Fig. 8 is an averaged line profile taken across middle third of this interface. It is immediately apparent that the profile is diffuse, dropping from 86% to 7% (the author's own criterion for edge resolution), not in 1 atomic plane as they claim in the paper, but rather over 3-4 planes. The derived EELS profile cannot be sharper than this elastic profile as it relies on the same probe shape. This should have ruled out any possibility for atomic resolution EELS, yet as seen in Suppl. Figs 9 and 10, the authors were still showing atomically abrupt EELS data. It is worth noting that in the second revision of the paper, the authors replaced without mention this image (Fig 4b of ref([3])) with the image now found in the published version of the Nature paper. When questioned by the referee, the authors denied this was a different image, even though the scan noise shows it was recorded in a different direction. They did, however, acknowledge that the unusually sharp interface that now appeared, and also ran halfway through the middle of some of the interface atoms, was an artifact and not related to the interface structure. Why they then knowingly chose to leave in a flawed image that artificially sharpened the interface, rather than restore an artifact-free one, is puzzling. Despite this, a line profile through this new image (Suppl. Figs 11&12) again shows a diffuse interface profile, spreading over 4-5 layers. Once again atomic-resolution EELS would have been impossible.

We now turn finally to the assurances the authors gave to Nature as reported in the accompanying editorial that "the original data, if consistently analyzed as intended, would still have supported the central thesis of the paper". In the 2$^{nd}$ and 3$^{rd}$ version of the Nature paper, the authors argue that a least squares fit to the reference spectrum of their Fig 4a would yield the EELS line profile shown in Fig 4b. It is Fig 4b that provides the edge profile evidence for atomic resolution. In the original version this figure was derived from a "jump-ratio" analysis. By overlaying the curves from the 2 versions some troubling details become apparent (Supp. Fig. 13). Note that only points 4 and 5 are different. This is disturbing as these are the 2 points on either side of the critical interface.

If Nature Fig 4b and MSA Fig 4c[3] were processed differently, the noise should be different and all 7 points should be different. (After normalizing one point to match, 6 of the 7 points should be different). If spectra 1-4 in Nature Fig 4b were processed the same as in MSA, then the first 4 points in both figures should match, but the last 3 should be different. Instead only points 4&5, the two spectra closest to the interface, and most critical to their claim, change. This cannot be explained by the original paper or the correction.

It is still possible to attempt a consistent analysis of the original data in the manner described, but not followed, by the paper, in order to see if the central thesis would still be supported. The digitized data from Suppl. Figs. 3 and 4 was quantified using a least squares fit to a digitized version of their reference spectrum (Nature Fig. 4a) and a constant offset to account for the offsetting of the spectra in the plot (Suppl. Fig 14). The two fitting parameters are a scale factor for the reference curve and the height of the background offset. Fitting to both the original data (MSA) and the published data (Nature, final version) shows that points 1-4 from the two sets essentially lie on top of each other, contrary to the authors' analysis (Supp. Fig 15 & 16). More importantly, the MSA line profile is no longer atomically abrupt, stretching over at least 2 atomic planes (i.e more than 6 Angstroms). Given the doubtful provenance of curves 5'-7' of the Nature paper, we return to the original data of the MSA paper. Suppl. Fig. 17 shows that this data is in good agreement with the elastic line profile through the annular dark field image of Fig 3a. Both profiles show a diffuse interface that is 2-5 atomic layers wide. This fails to meet the authors' own definition of atomically abrupt and contradicts their recent assurances.

In summary, the original data did not support their central claim of atomic resolution EELS; neither does the reanalyzed data. The alterations to the data set and incorrect claims made during the refereeing process and in the corrigendum further undermine confidence in their data. At this point, it would seem that the only reasonable course of action is a retraction of the paper.

The changing EELS data of Fig 3a in the 3 versions submitted to Nature by Browning et al.

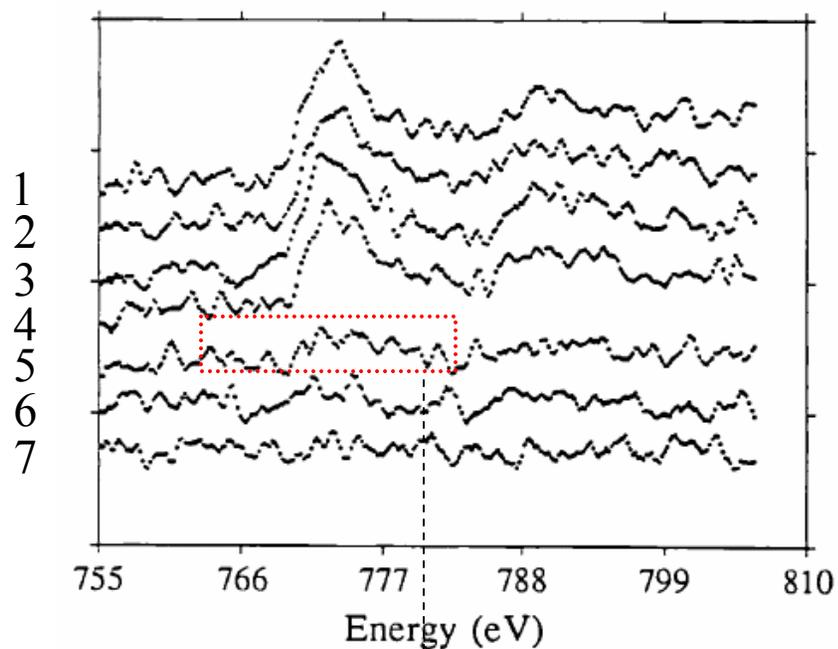

Fig 3a, 1st version

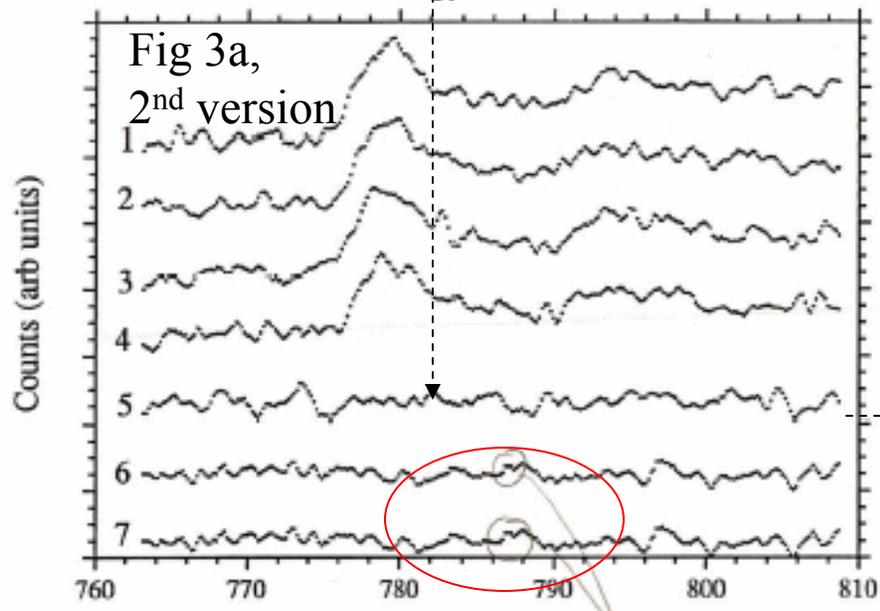

Fig 3a, 2nd version

Supplementary Figure 1    Duplicate curves

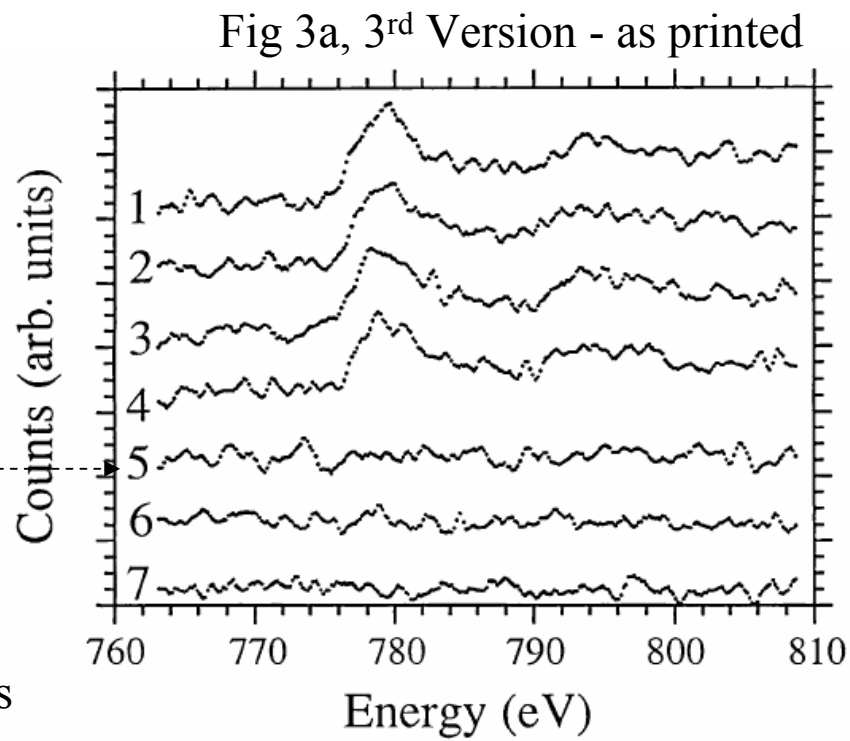

Fig 3a, 3rd Version - as printed

EELS data of MSA Fig 4a after the constant offset is removed

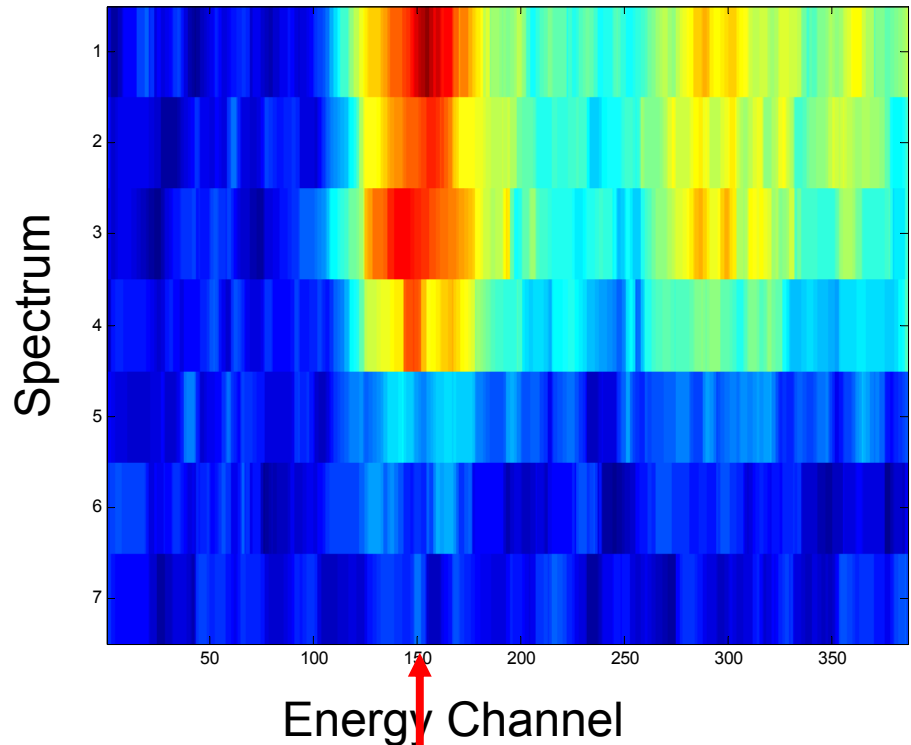

EELS data of Nature Fig 3b after the constant offset is removed

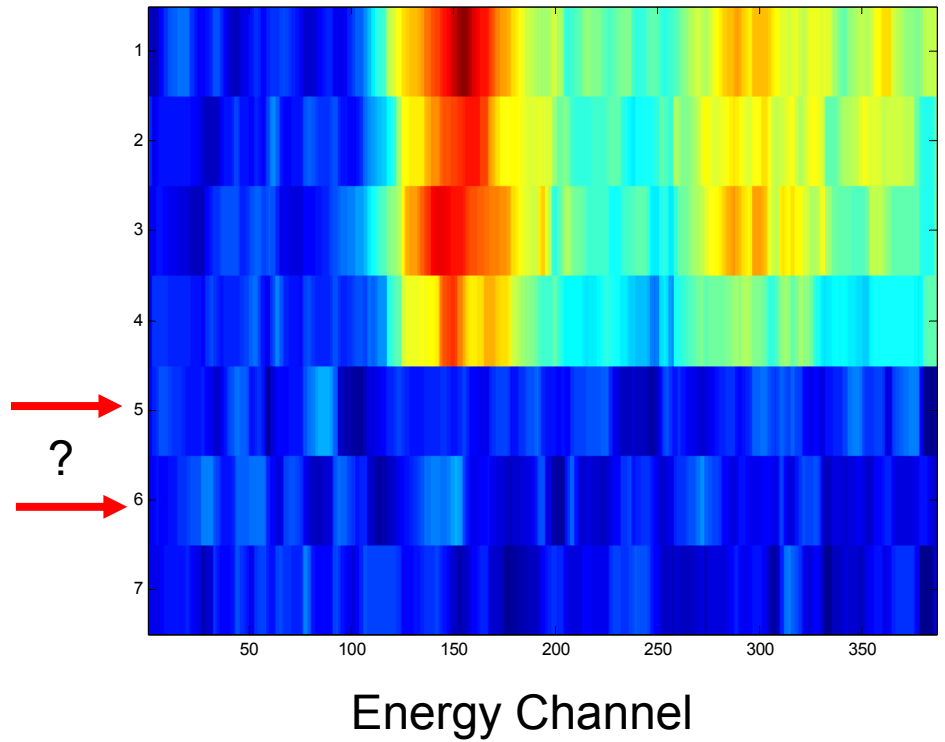

A significant Co tail is present in the MSA paper, but has been removed from the Nature data.

Spectrum 4 is also much lower in intensity than 1-3 (not reflected in "processed" Nature curve of Fig 4b)

Supplementary Figure 2

Nature Fig 3b (print version) - Scanned and Digitized

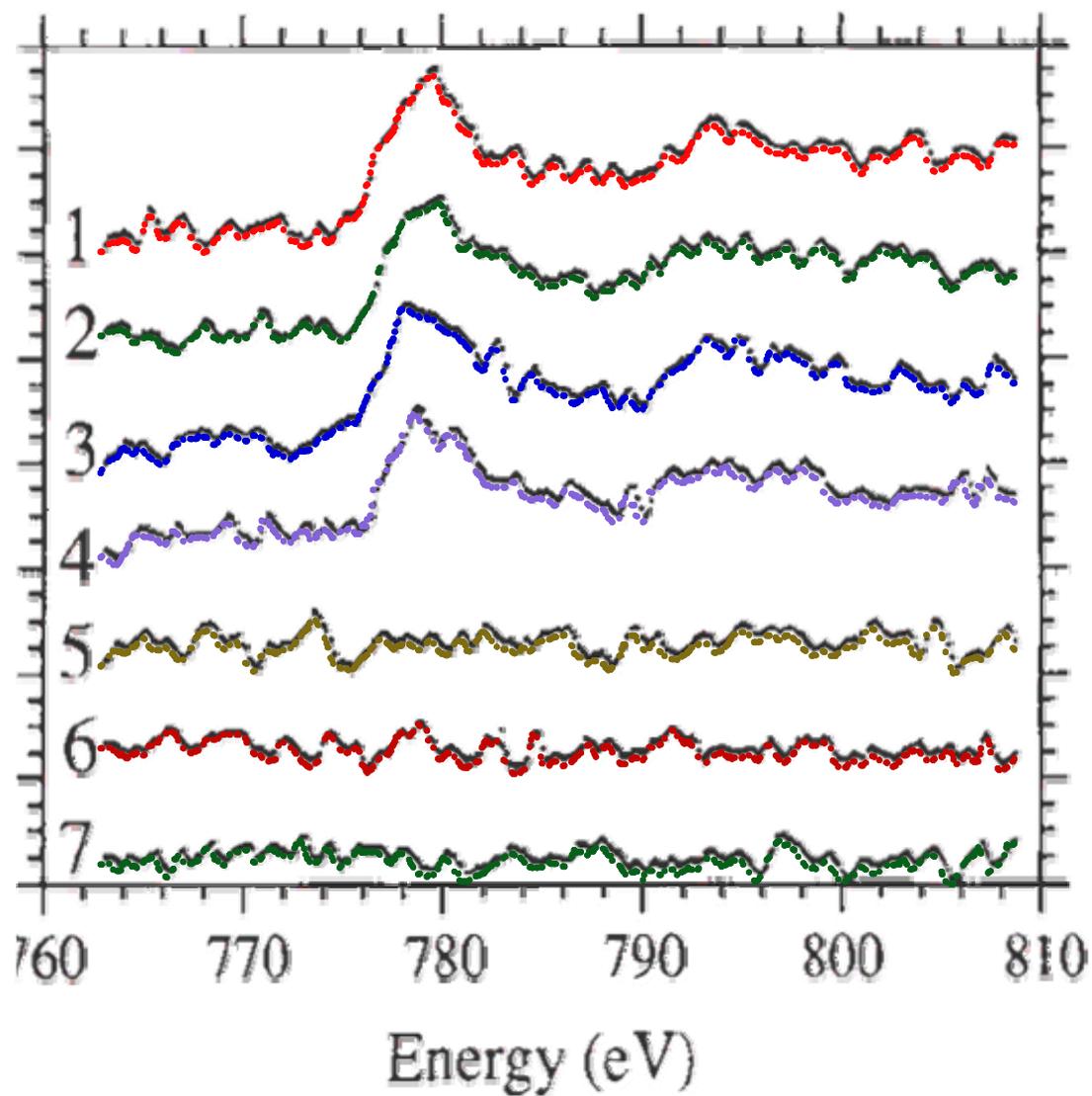

Supplementary Figure 3

Browning et al, MSA 93 - Scanned and Digitized

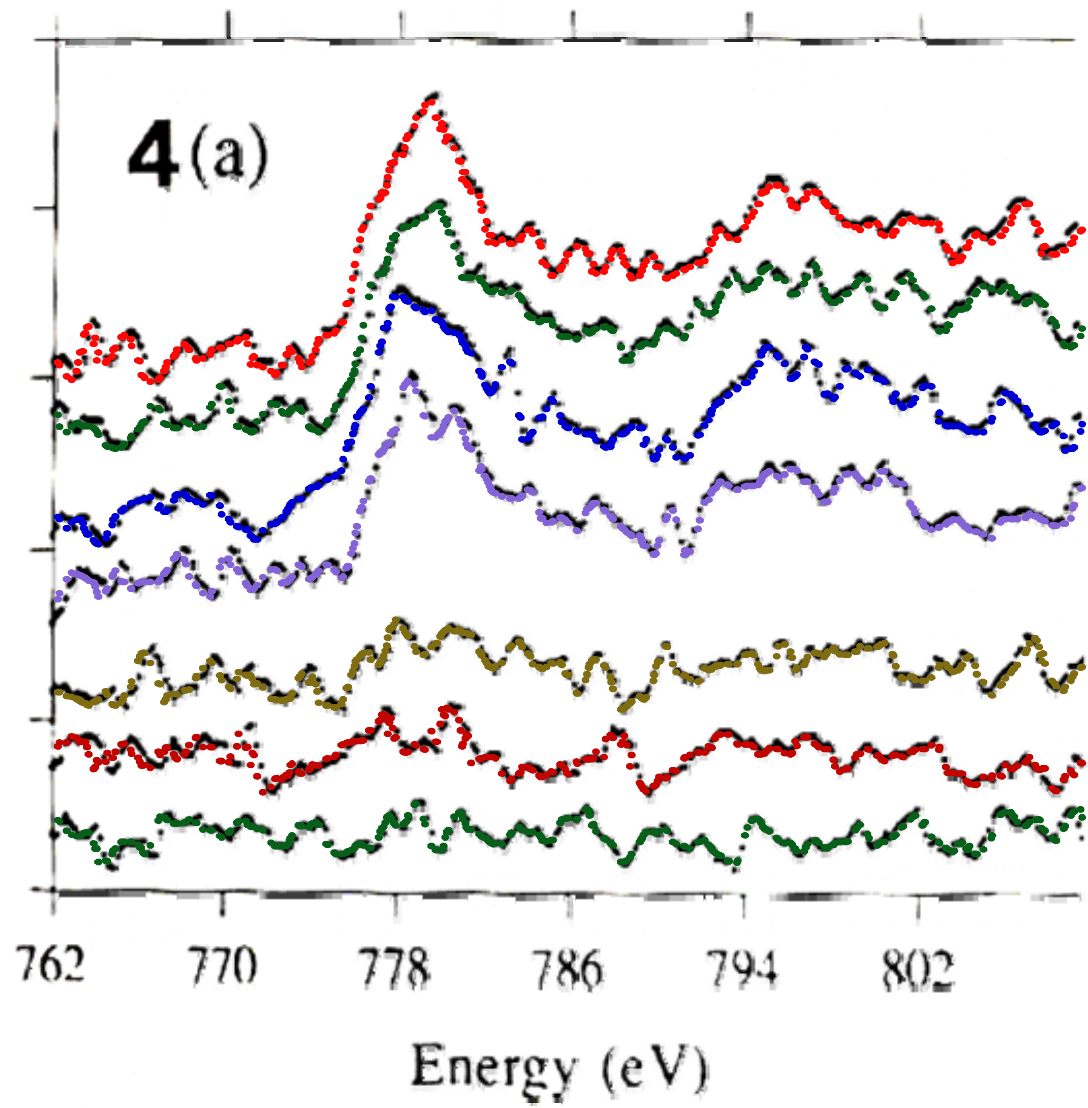

Supplementary Figure 4

Data from MSA '93 Fig 4a rescaled to the same axis Nature '93 Fig 3b, and plotted against each to show the good 1:1 match for spectra 1-4, and scatter from the poor match for 5-7.

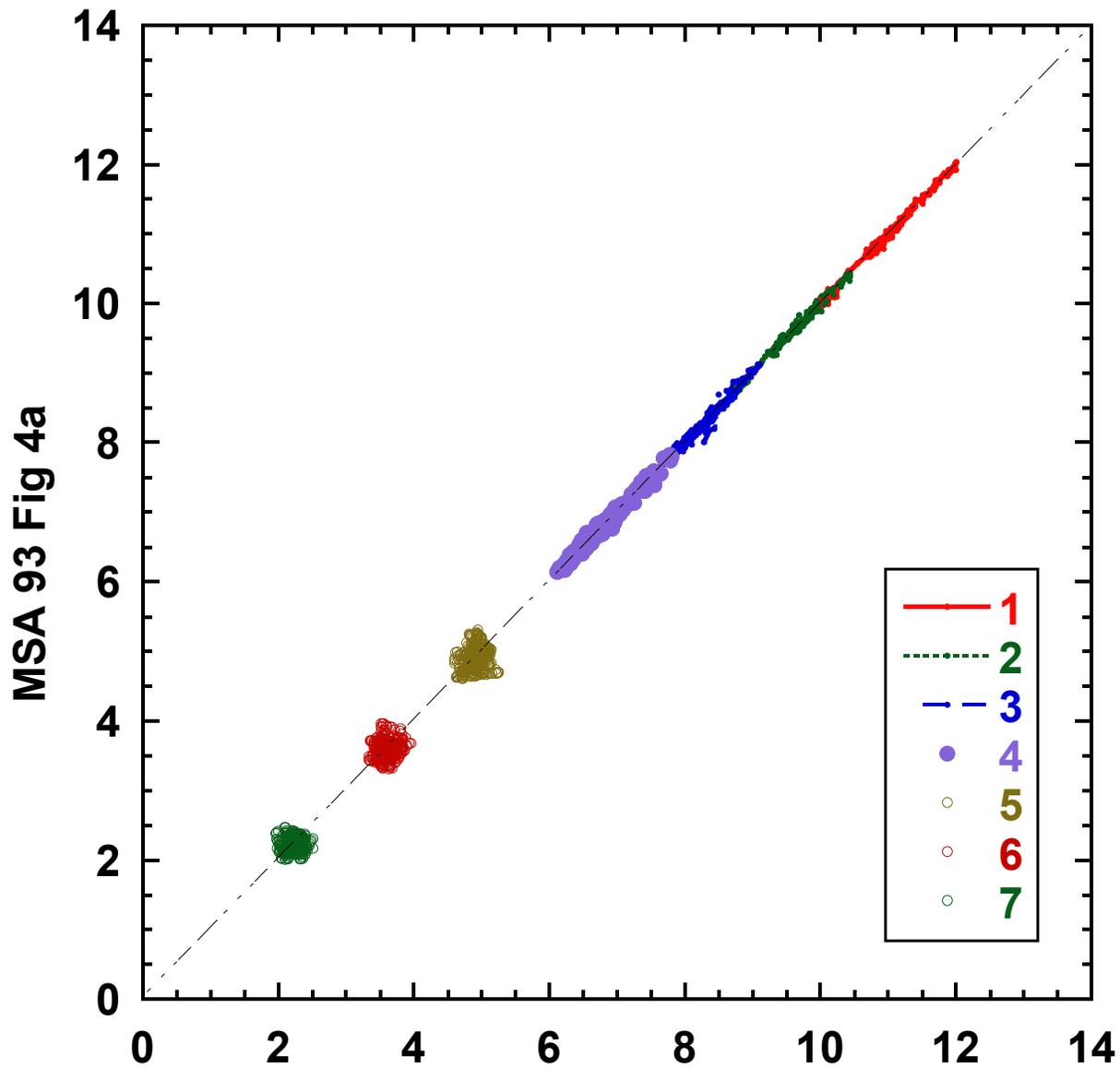

Supplementary Figure 5

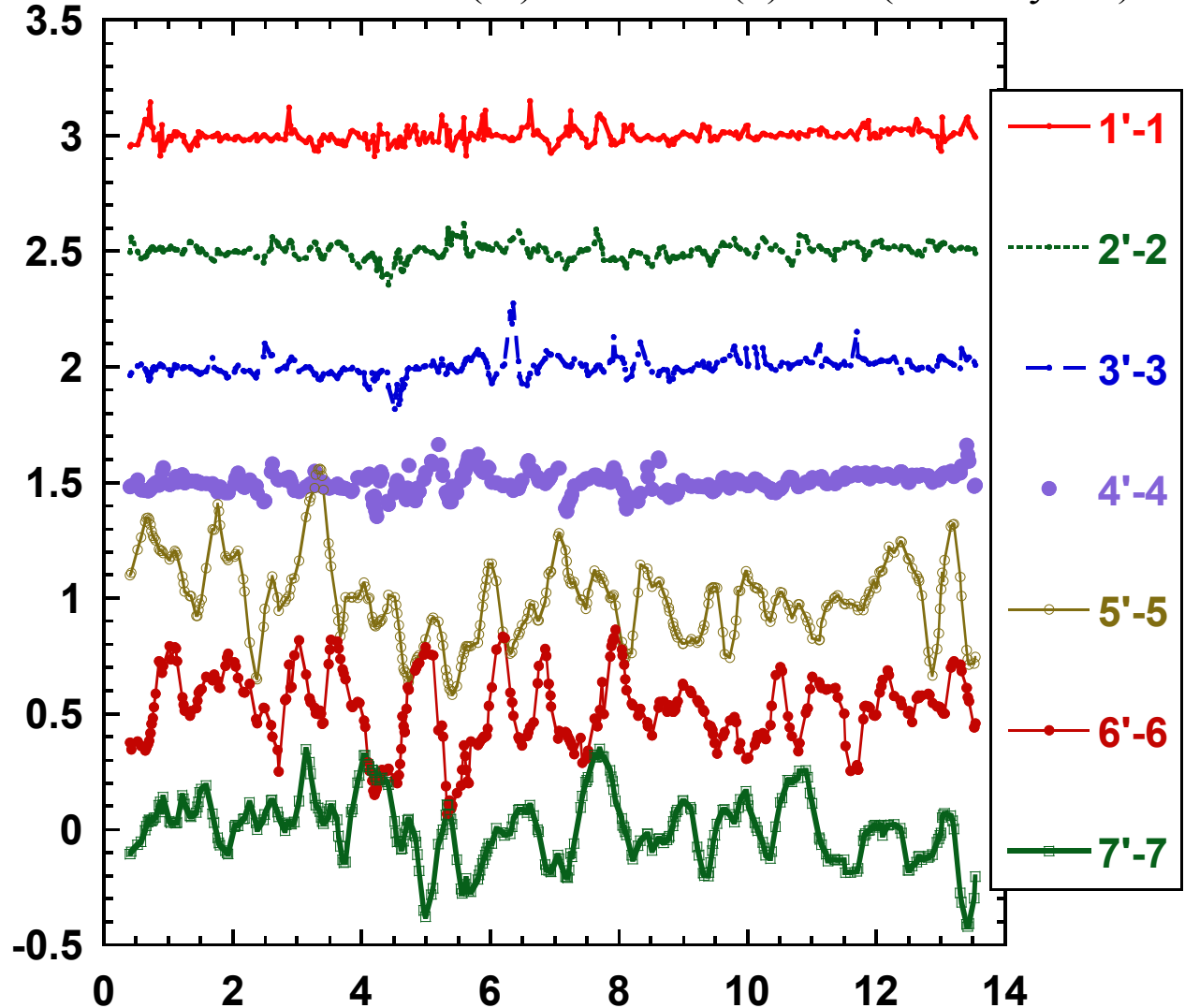

The large residuals in 5-7 should all have had the same shape if the curves 5-7 were the same in the Nature and MSA papers, and the same reference curve was subtracted from the Nature curves. The lack of any correlation suggests these are **different data sets, contradicting the claim made by the authors both in the corrigendum and in 1993.**

Supplementary Figure 6

Browning, N.D., M.F. Chisholm, and S.J. Pennycook. *Atomic-resolution chemical analysis using a scanning transmission electron microscope*. in *51st Ann. Proc. Microsc. Soc. Am.* 1993: San Francisco Press.

Figure 4b

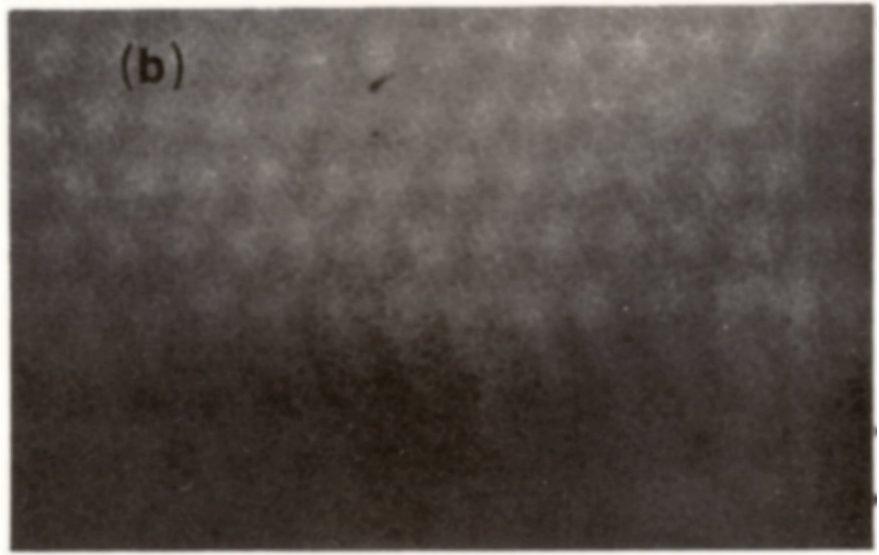

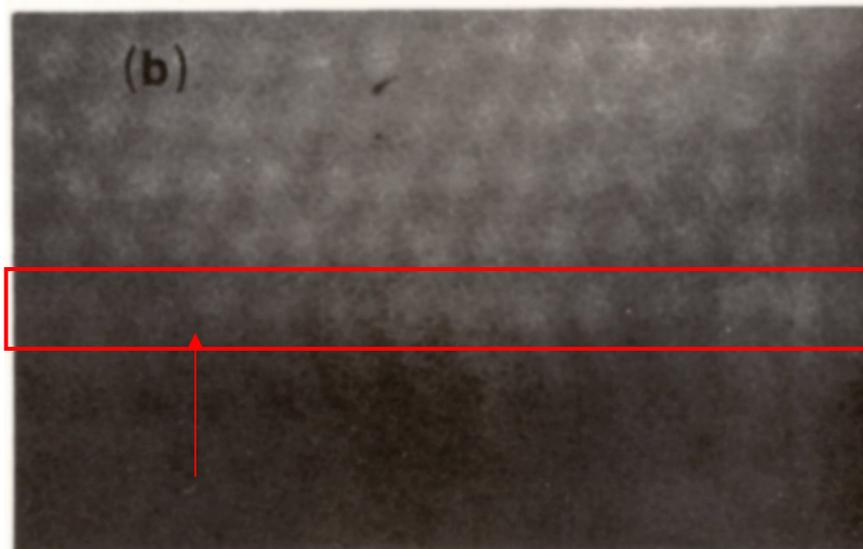

Interface plane is rough and contains a step – red arrow

Supplementary Figure 7

Intensity Profile averaged plane-by-plane across the center of Fig 4b of MSA 93 compared to the EELS profile (red) in Fig 4c of MSA93 (identical to Nature V1)

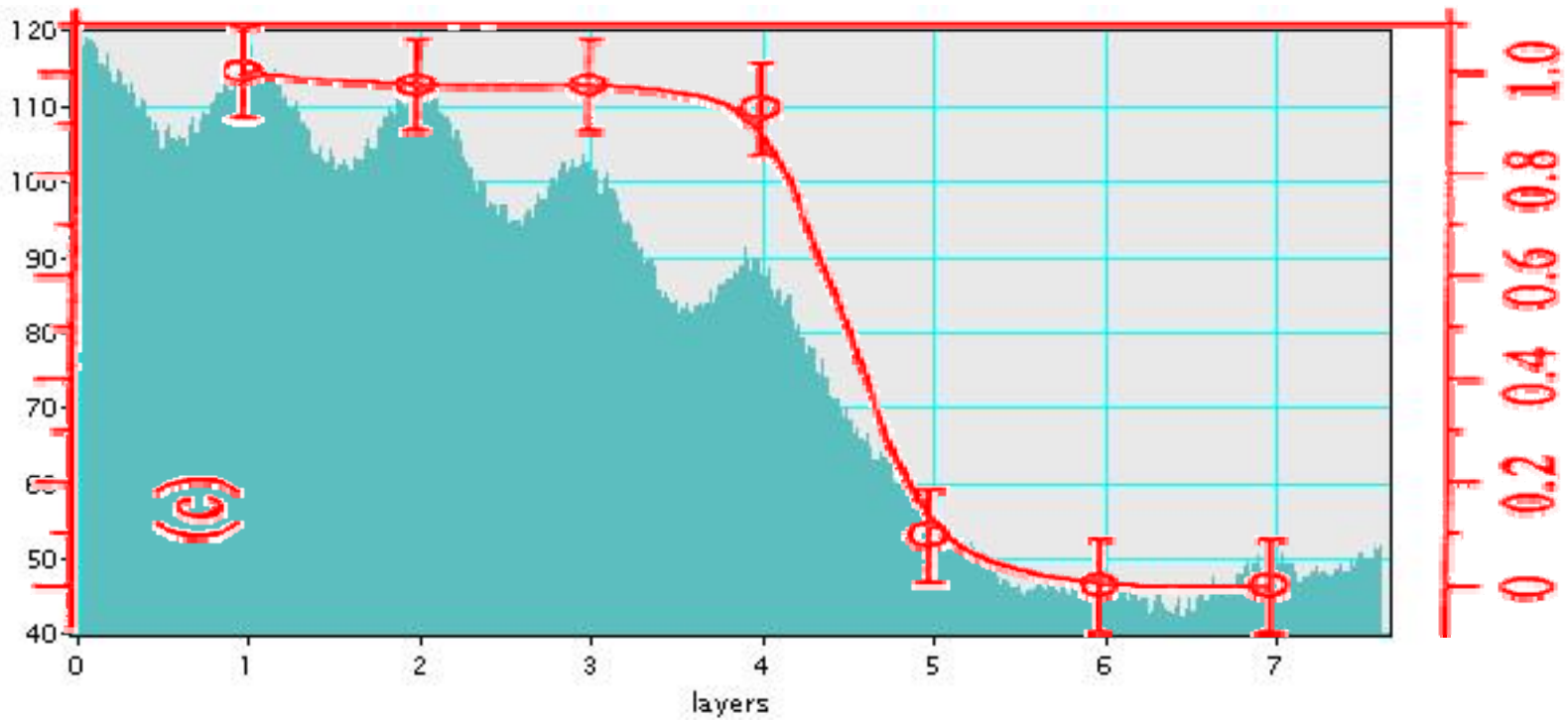

The steady decrease of intensity on columns 2,3&4 are symptomatic of probe tails due to an improperly large probe-forming aperture. If atomic-resolution EELS had really been achieved, the EELS profile of Fig 4c should have followed this shape.

The EELS profile is sharper than the Elastic Image profile. This is unphysical as the image profile should describe any probe broadening and beam spreading. The EELS profile can only be broader than the elastic signal.

Supplementary Figure 8

Intensity Profile averaged plane-by-plane across Fig 4b of MSA 93 compared to the EELS profile in Fig 4 of Nature V1. o –expt, x - theory

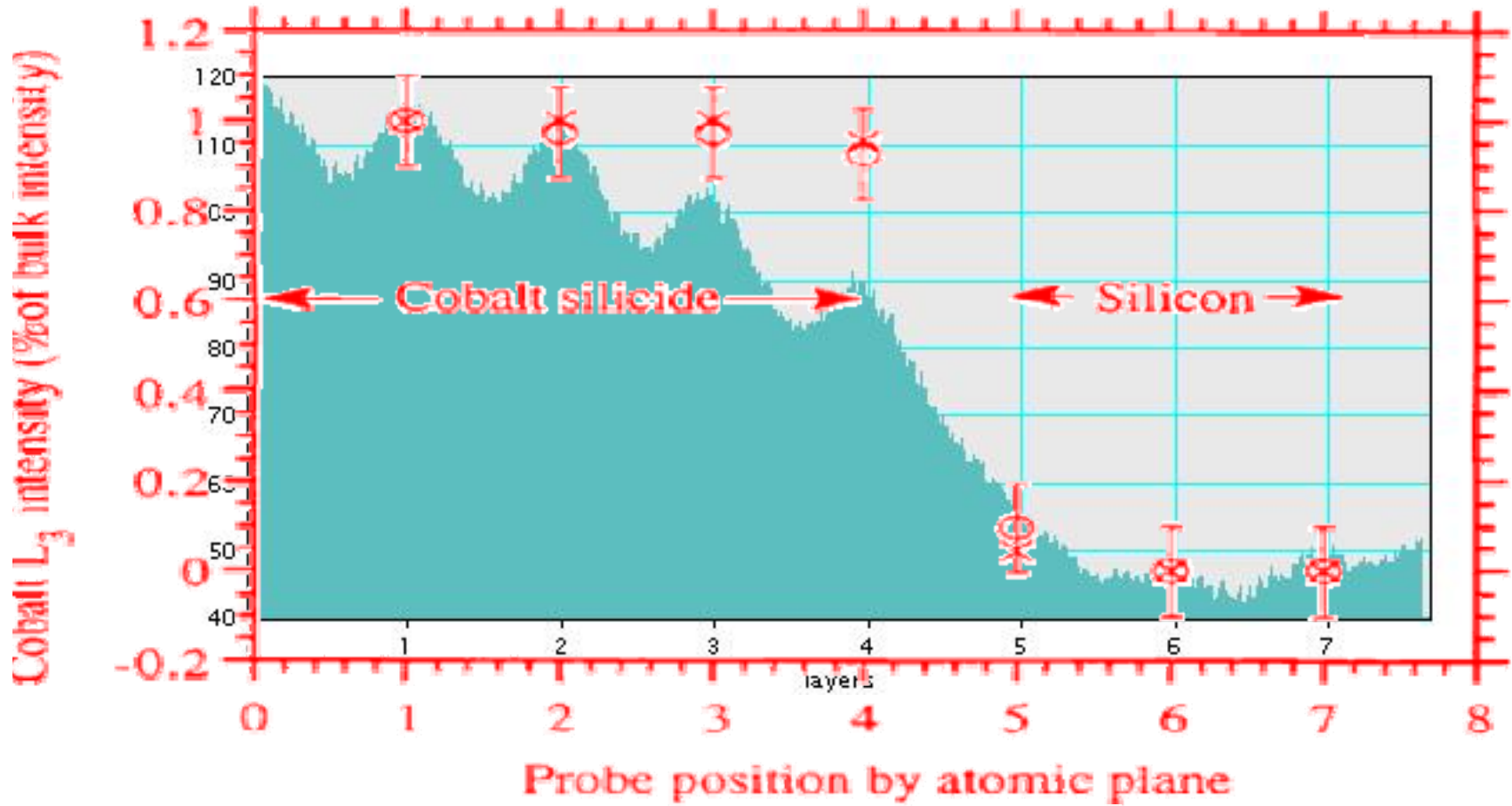

The EELS profile is in remarkably good agreement with the theory profile. However, the theory describes a profile for an ideal interface. The real interface profile is given by the elastic image profile of MSA Fib 4b (blue curve). The sharp EELS could not have come from this interface.

Supplementary Figure 9

Intensity Profile averaged plane-by-plane from a **wider region** across Fig 4b of MSA 93 compared to the EELS profile in Fig 4 of Nature V1. o –expt, x - theory

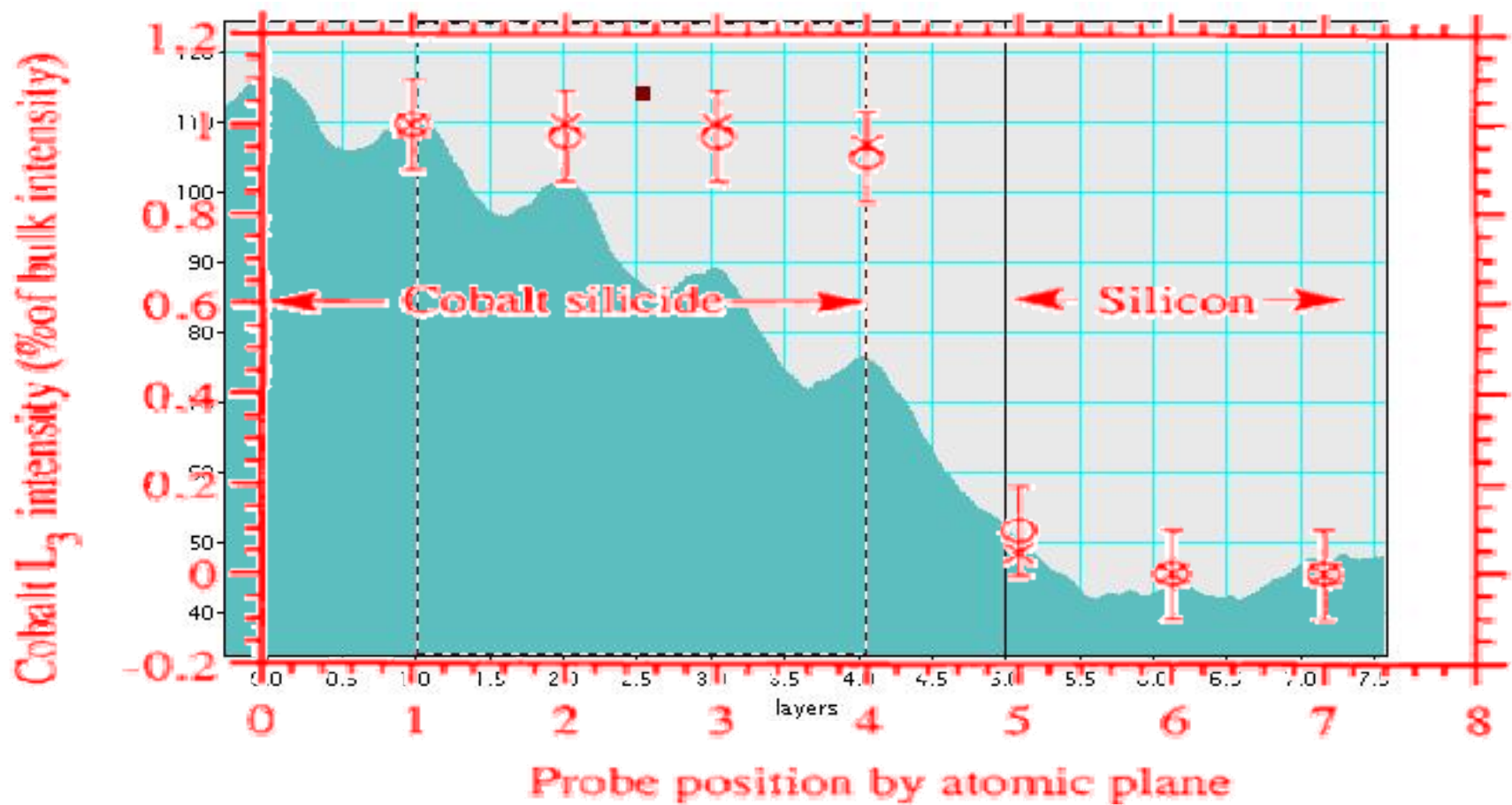

Supplementary Figure 10

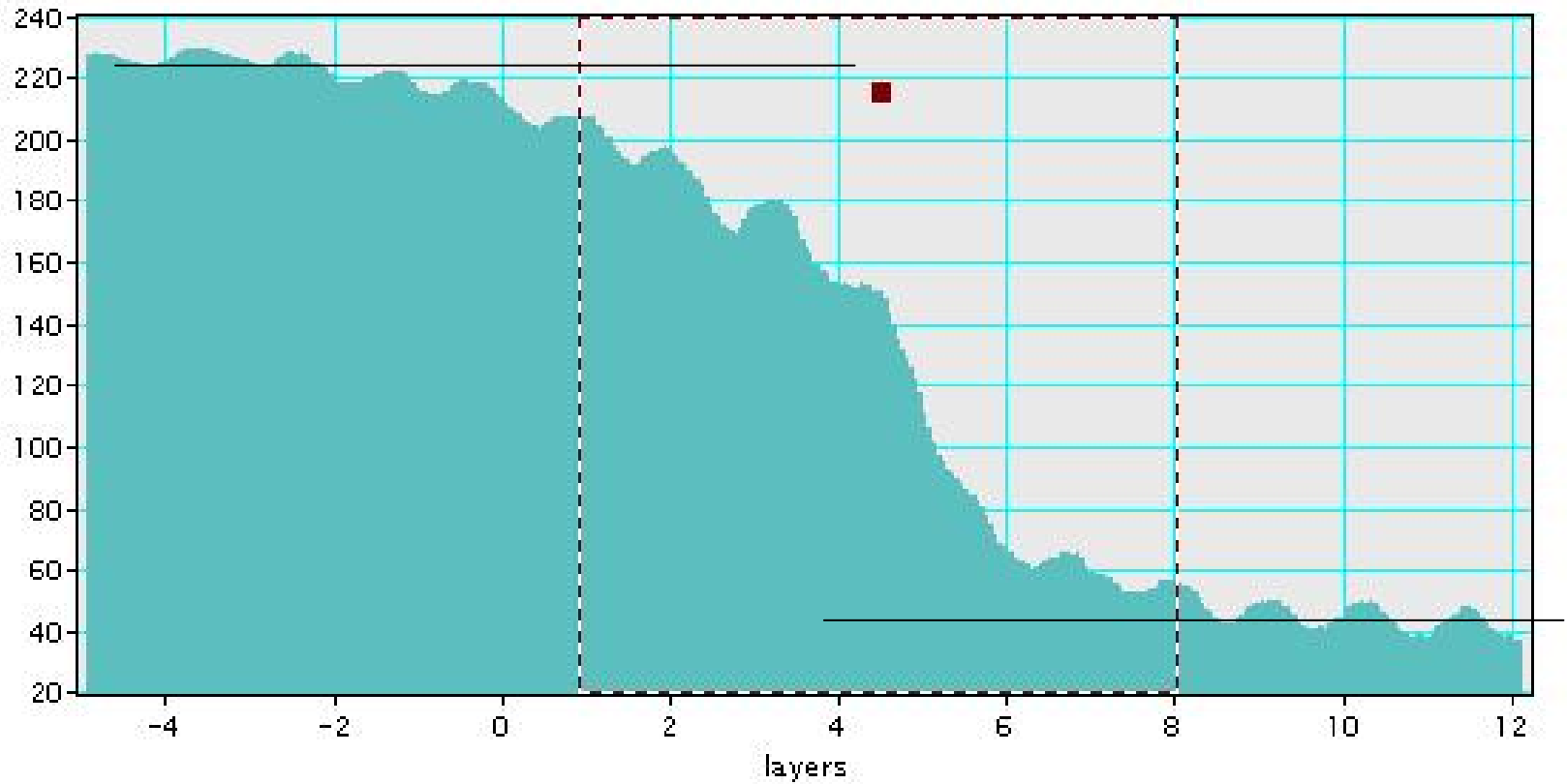

Averaged line profile through Nature Fig 3a. The horizontal black lines show the bulk levels for the $CoSi_2$ and Si sides of the interface. The interface profile is extended over 5 layers, despite an unphysical midway through the atoms in layer 4. Any EELS profile cannot be more abrupt than this profile.

Supplementary Figure 11

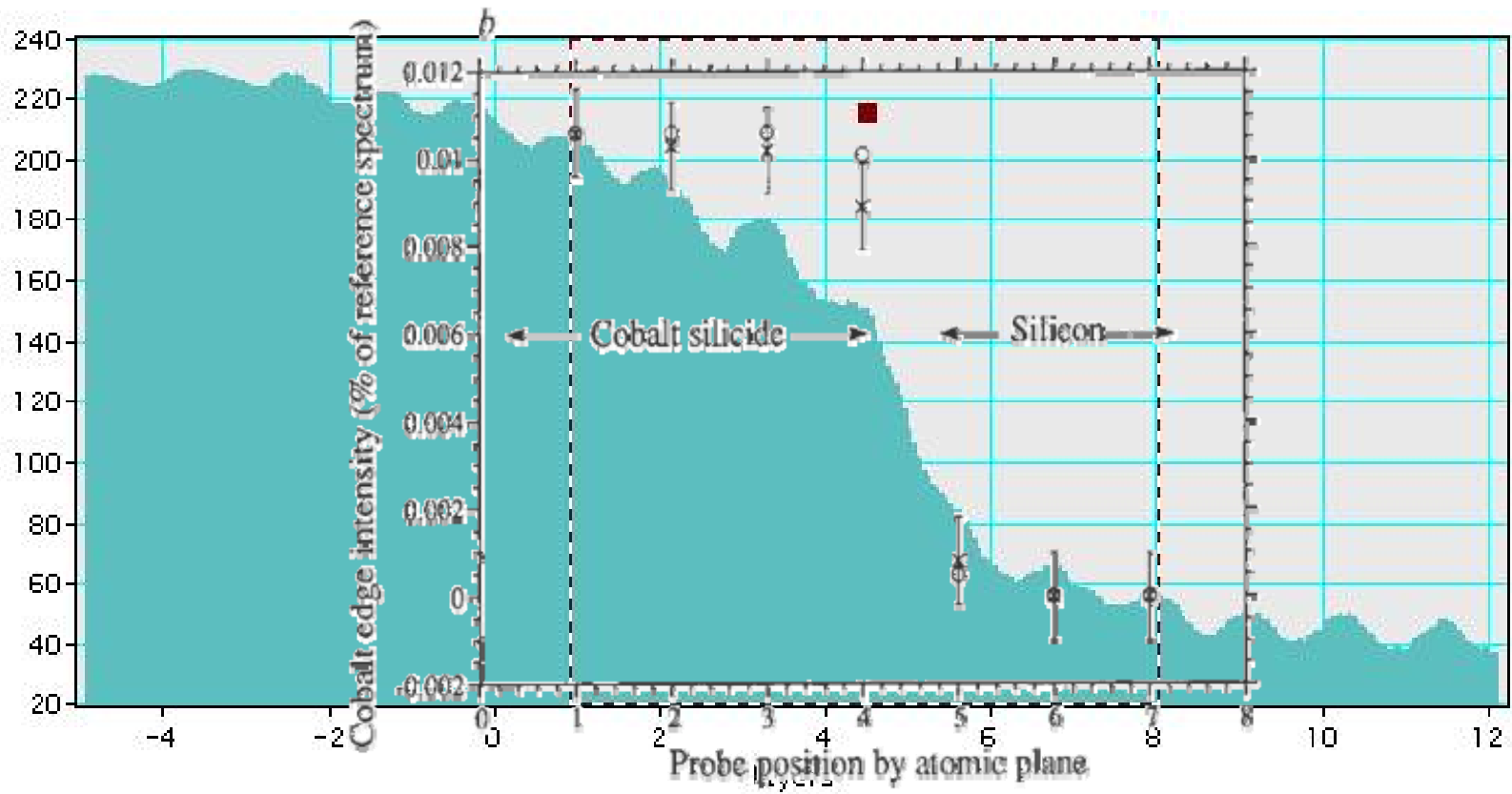

Averaged line profile through Nature Fig 3a and EELS data from Nature Fig 4b. The theory points (o) are supposed to follow the line profile of fig 3a if fig 3a were really atomically abrupt. While the "experiment"-(x) from 4b matches the theory from 4b, neither matches with the interface profile from the atomic-resolution image, fig 3a.

Supplementary Figure 12

Nature 93 Fig 4b (V3) vs MSA 93 fig 4c

Data from MSA 93 fig 4c is identical to that submitted to Nature as the 1st version
2nd and 3rd versions of Nature fig 4b are identical to the print version.

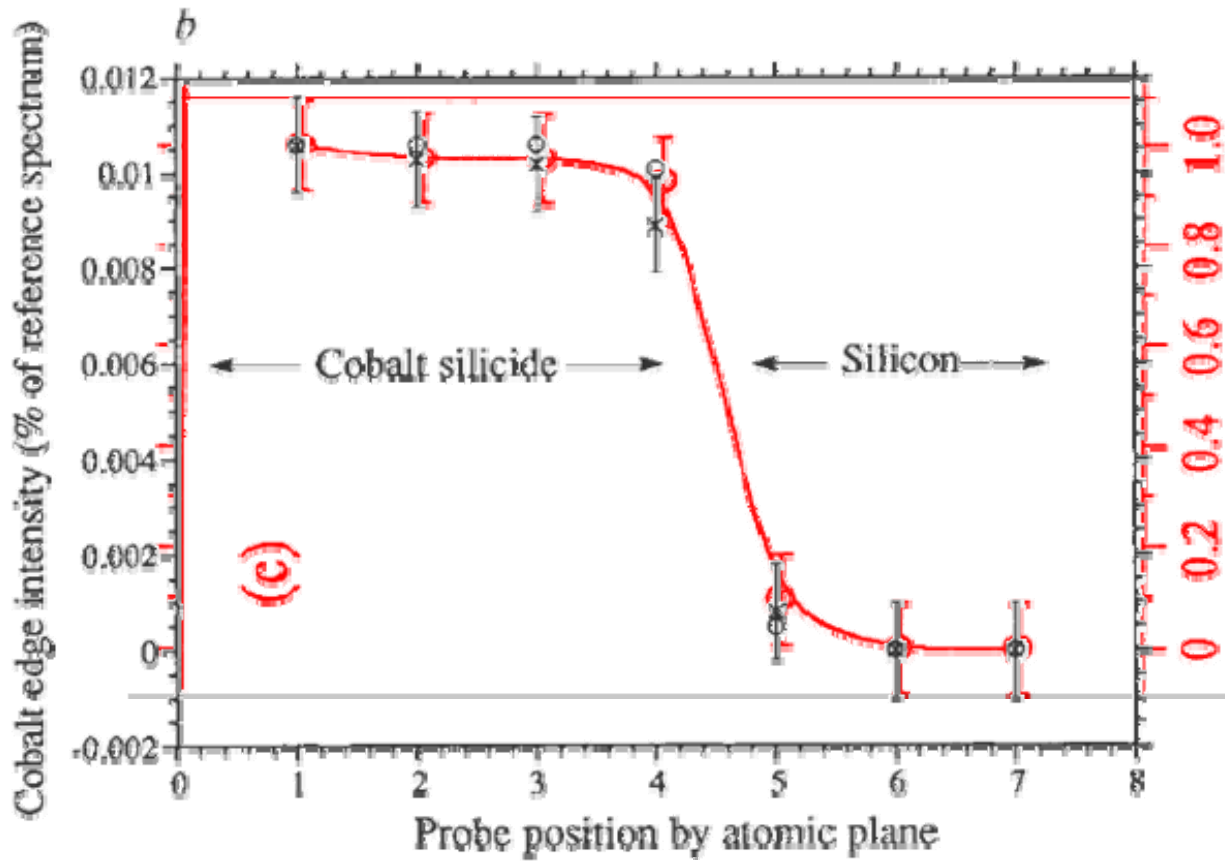

MSA 93
o expt,

Nature 93 (in print)
x expt,
o theory

Note difference
with Supp.Fig.10.

Note that points 4,5 are the only different ones.

Supplementary Figure 13

# Least-Squares Fit (red dotted line) to MSA 93 and Nat 93 digitized data

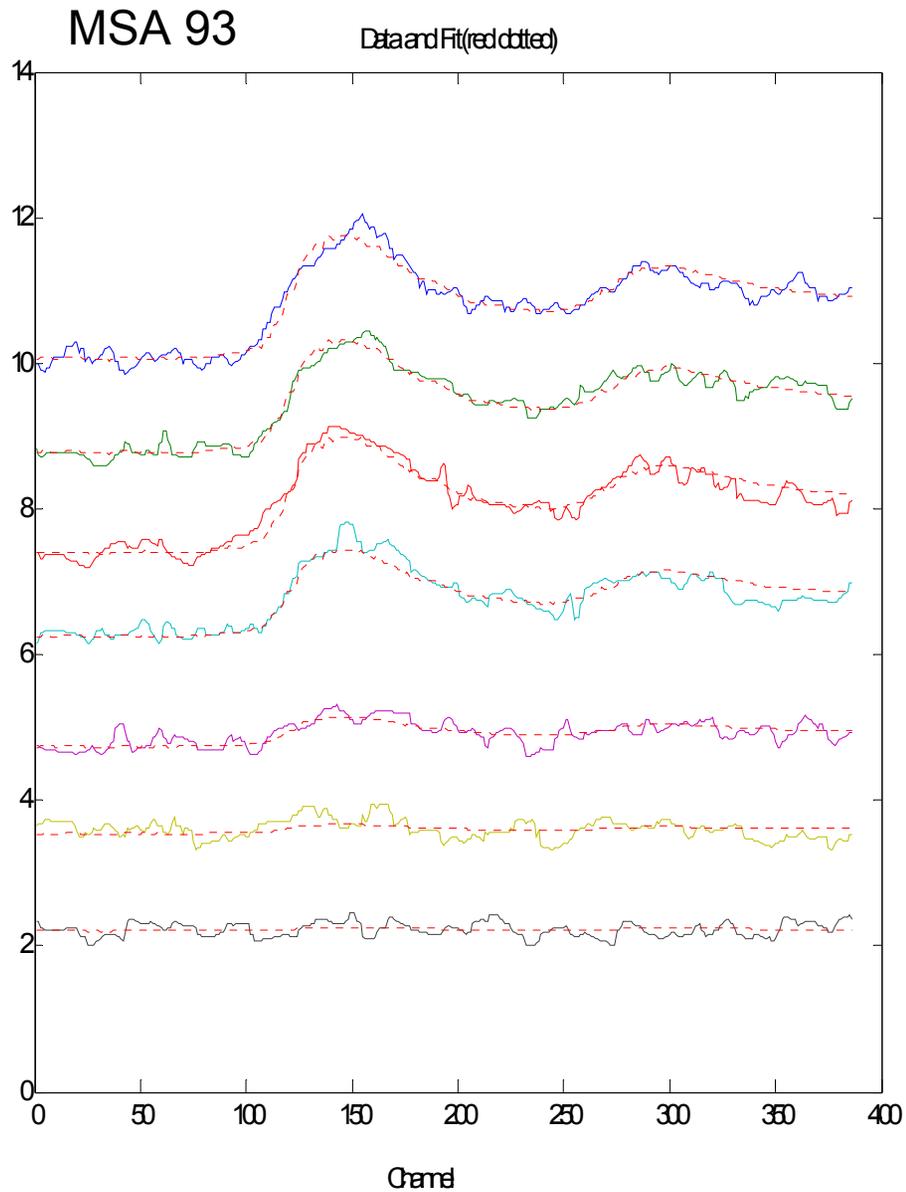
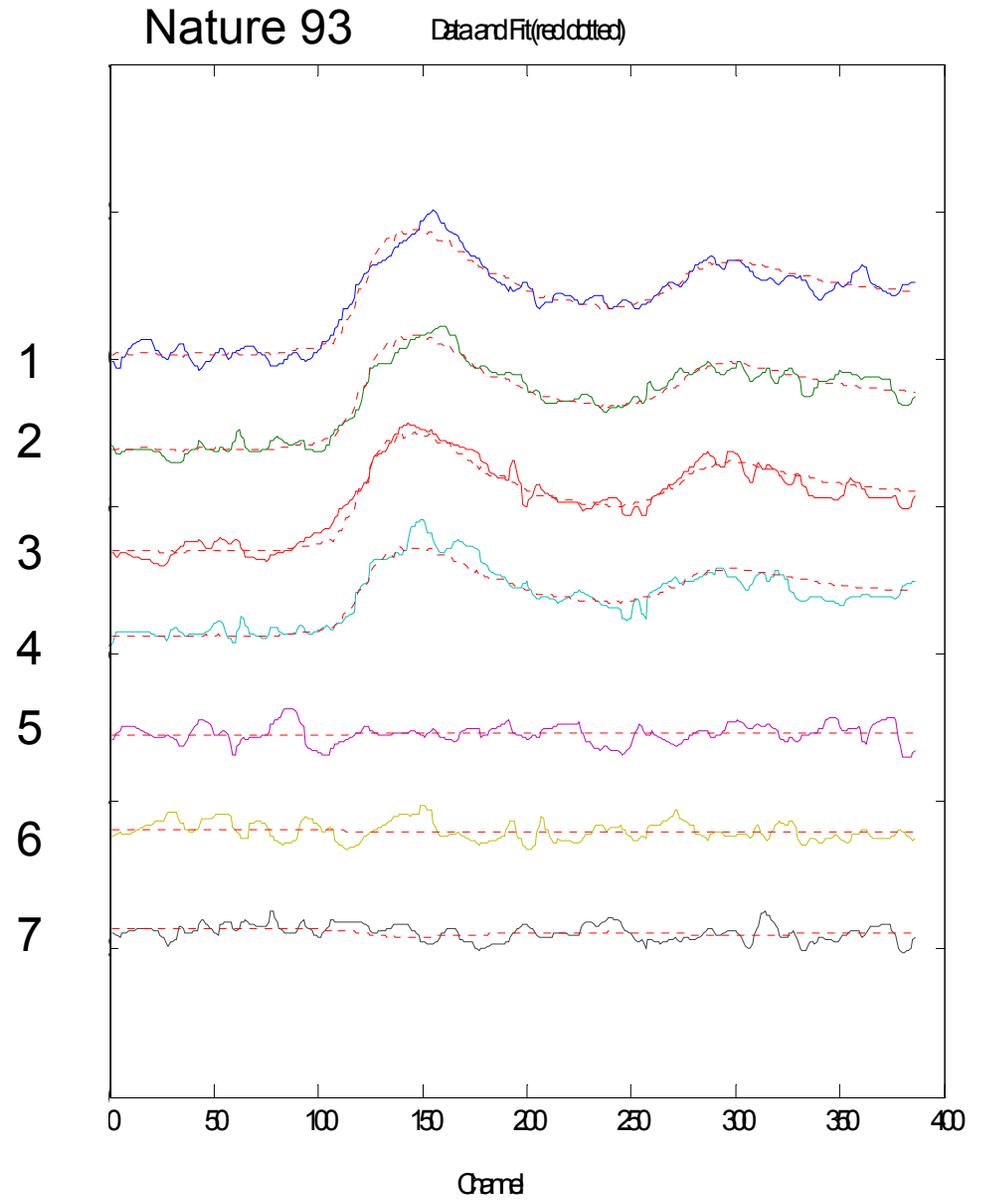

Supplementary Figure 14

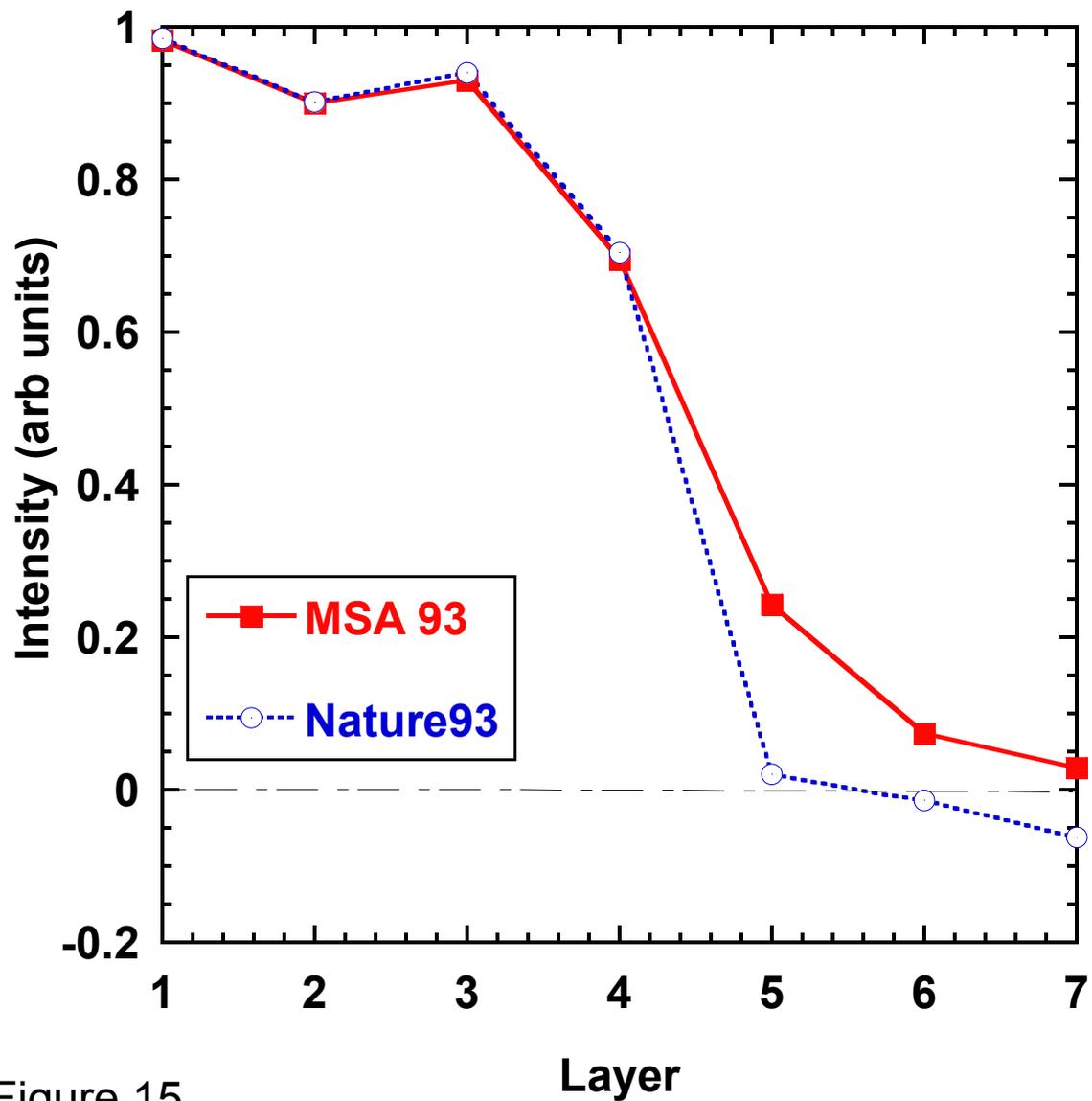

Supplementary Figure 15

Nature 93 Published version: x-expt   o-theory

Vs least squares fit (rescaled to match 1st point)

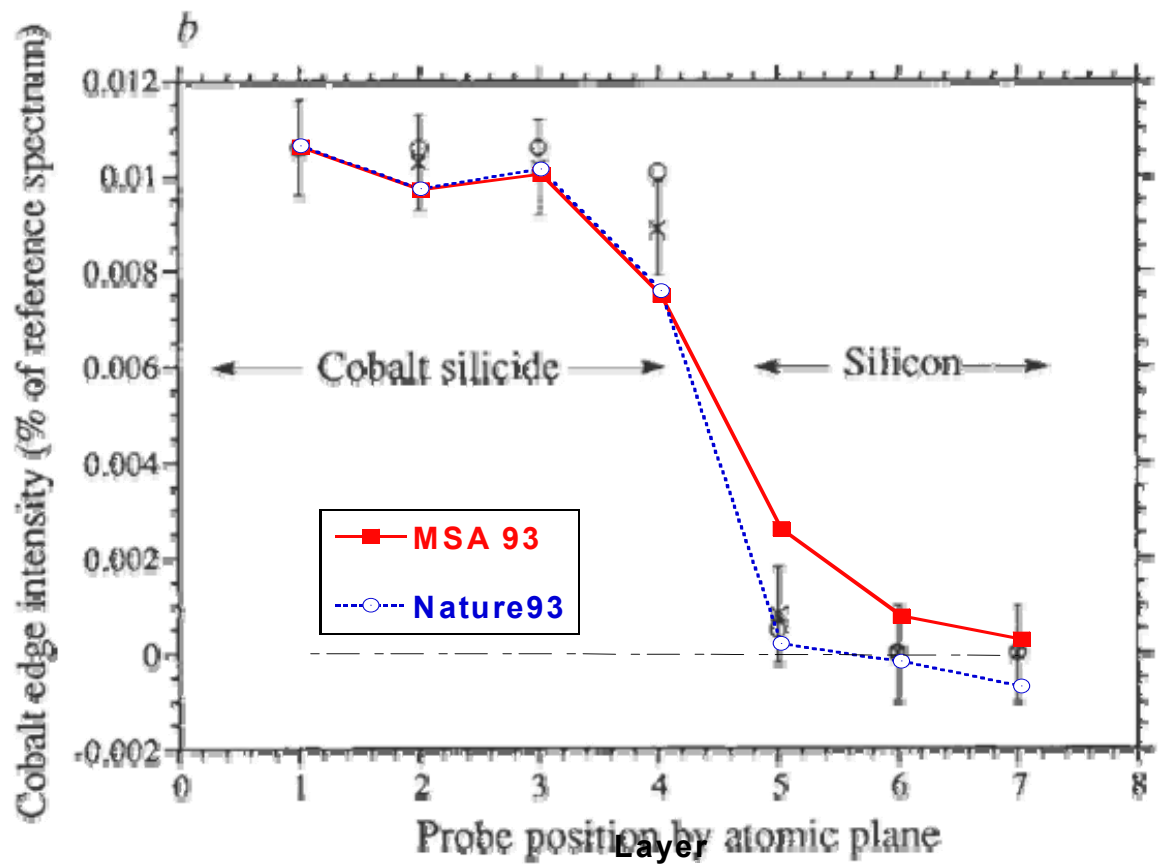

Supplementary Figure 16

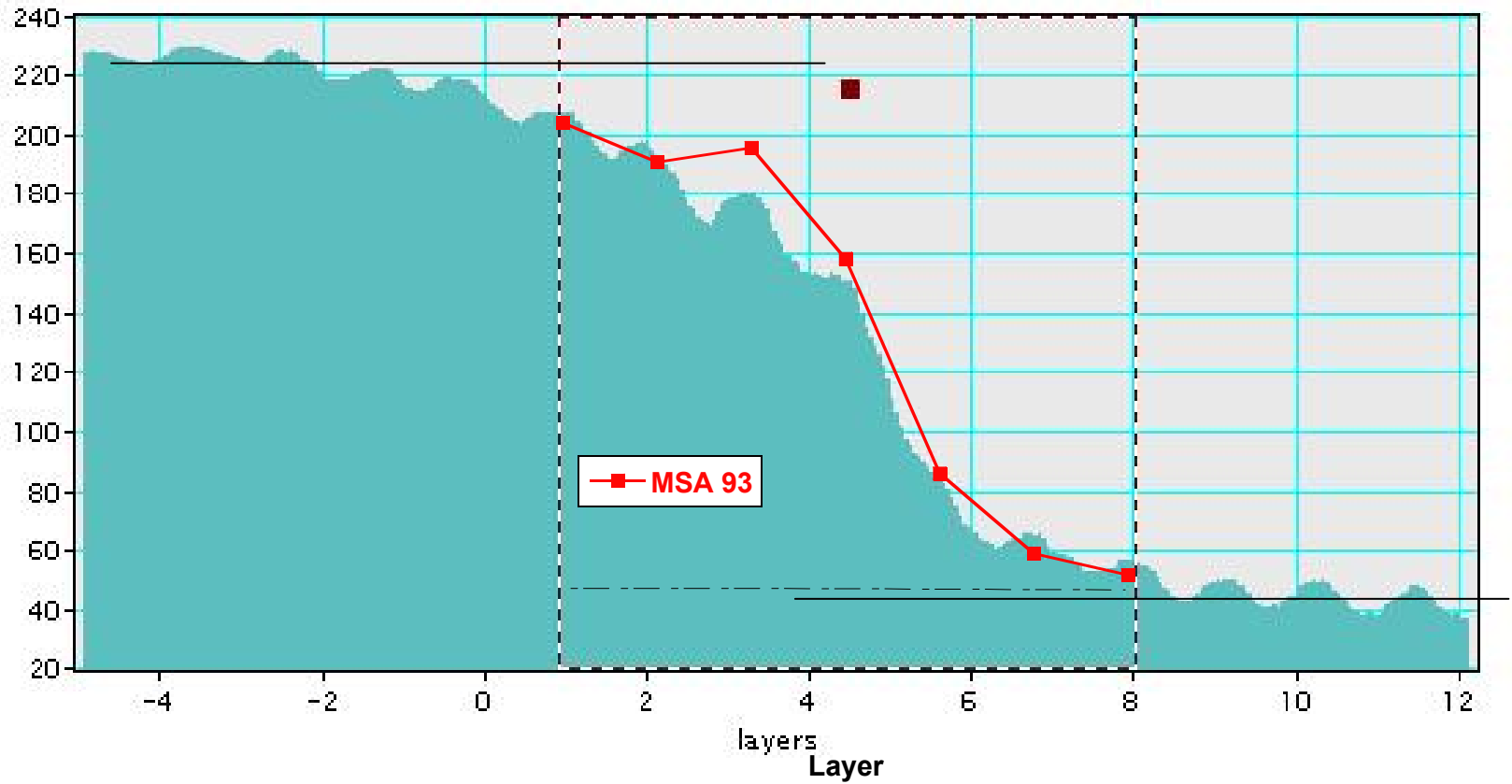

Averaged line profile through Nature Fig 3a. The horizontal black lines show the bulk levels for the $CoSi_2$ and Si sides of the interface. The interface profile is extended over 5 layers, despite an unphysical midway through the atoms in layer 4. The red curve is the least squares fit to the MSA 93 EELS data. It is in good agreement with the elastic profile, both showing a very extended interface width.

Supplementary Figure 17